# Graphene Segregated on Ni surfaces and Transferred to Insulators

Qingkai Yu

*Center for Advanced Materials, Electrical and Computer Engineering, University of Houston, Houston, Texas 77204*

Jie Lian

*Department of Mechanical, Aerospace and Nuclear Engineering, Rensselaer Polytechnic Institute, Troy, New York 12180*

Sujitra Siriponglert

*Center for Advanced Materials, University of Houston, Houston, Texas 77204*

Hao Li

*Department of Mechanical and Aerospace Engineering, University of Missouri, Columbia, Missouri 65211*

Yong P. Chen

*Department of Physics and Birck Nanotechnology Center, Purdue University, West Lafayette, Indiana 47907*

Shin-Shem Pei

*Center for Advanced Materials, Electrical and Computer Engineering, University of Houston, Houston, Texas 77204*

**ABSTRACT**

**We report a surface segregation approach to synthesize high quality graphenes on Ni under ambient pressure. Graphenes were segregated from Ni surfaces by carbon dissolving at high temperature and cooling down with various cooling rates. Different segregation behaviors were identified, allowing us to control the thickness and defects of graphene films. Electron microscopy and Raman spectroscopy studies indicated that these graphenes have high quality crystalline structure and controllable thickness. Graphenes were transferred to insulating substrates by wet etching and were found to maintain their high quality.**

Graphene,[1] the 2D counterpart of 3D graphite, has attracted vast interests in solid-state physics, materials science and nano-electronics, since it was discovered in 2004 as the first free-standing 2D crystal. Graphene is considered as a promising electronic material in post-silicon electronics. However, large-scale synthesis of high quality graphene represents the bottleneck for the next generation graphene devices. Existing routes for graphene synthesis include mechanical exfoliation of highly-ordered pyrolytic graphite (HOPG),[2, 3] eliminating Si from single crystal SiC surface,[4, 5] depositing graphene at the surface of single crystal[6] or polycrystalline metals[7] and various wet-chemistry based appraches.[8-10] However, up to now no methods have delivered high quality graphene with large area required for application as a practical electronic material.

Graphite segregation at surfaces and grain boundaries of metals has been studied for a long time.[11] Graphite with low defects density can segregate from metals and metal carbides.[6, 12] However, the control of segregation behavior of graphite to produce graphene is much less studied. Here, we demonstrate the synthesis of several layers of graphene on Ni substrates in large area by surface segregation. Controlling synthesis parameters, especially the cooling rate, is critical to produce thin graphene film (<10 layers). We also demonstrate the transfer of graphene from metal substrates to insulating substrates. The graphenes maintain their high quality after transfer, as confirmed by Raman spectroscopy.

Graphene segregation by cooling is a non-equilibrium phenomenon. Non-equilibrium segregation in general involves the transport of vacancy-impurity (vacancy-carbon in our case) complexes to sinks, such as grain boundaries and surfaces during cooling, and is closely related to the cooling rate.[13] Our strategy is to control the amount of carbon segregated from metals by controlling the cooling rate, as illustrated in Fig. 1. In the first step, metal foils are placed in a chamber at high temperature with inert gas protection. In the second step (carbon dissolution), hydrocarbon gases are introduced to the chamber as the source of carbon. The hydrocarbon molecules decompose at the metal surface and diffuse into the metal. The concentration of carbon in the metal has an exponentially decreasing distribution from the surface into the bulk. This step is kept with a short time, generally 20 mins, to achieve the low carbon concentration. In the last step, samples are cooled down for carbon segregation. Different cooling rates lead to different segregation behaviors. Extremely fast cooling rate gives rise to a quench effect, in which the solute atoms lose the mobility before they diffuse. With a wide range of medium cooling rates, a finite amount of carbon can segregate at the surface. The extremely slow cooling rate allows carbon with enough time to diffuse into the metal body, so there will not be enough carbon segregation occurring at the surface.

Polycrystalline Ni foils with thickness of 0.5 mm and purity >99.99% from Alfa Aesar were cut into 5 mm×5 mm pieces, followed by a mechanical polish. Precursor gases are $CH_4:H_2:Ar=0.15:1:2$ with total gas flow rate of 315 sccm and pressure at 1 atm, with $H_2$ introduced one hour before the $CH_4$ and Ar. Carbon dissolution time is 20 mins at 1000 °C. Samples were cooled down by mechanically pushing the sample holder to lower temperature zones in the range of 30~500 °C in Ar atmosphere. Cooling rates were monitored by a thermal couple on the sample holder. Different cooling rates, corresponding to fast (20 °C/s), medium (10 °C/s) and slow (0.1 °C/s), were employed, and the structural characteristics of graphenes formed on Ni substrates were studied by transmission electron microscopy (TEM) and Raman spectroscopy (excited by an Argon laser operating at 514.5 nm).

TEM samples were prepared by peeling off graphene films in $HNO_3$ solution, followed by rinsing by deionized water. Graphenes float on the surface of water owing to its hydrophobic nature. The graphenes, found to be almost transparent, can nonetheless be distinguished from water by their different reflectivity. Copper grids with Famvar films were used to dredge up graphenes, which were then dried in air naturally. Edges of the graphene, as highlighted by red dash lines in Fig. 2a, have step features, which may be attributed to graphene cracking along certain crystalline directions. The selected area electron diffraction pattern (SAED) along [001] direction of graphene films clearly shows graphite lattice structure, and 3~4 layers of graphene were observed at the wrinkles and edges of graphene films as shown by the high resolution TEM image (HRTEM) (Fig. 2b).

Using Raman spectroscopy, we have characterized thoroughly the nature of the films and the numbers of graphene layers segregated on Ni substrates with different cooling rates (Fig. 3). Generally, four distinct patterns in the Raman spectrum can be used to characterize graphene:[14, 15] 1) The 2D band peak at ~2700 $cm^{-1}$ is symmetric for graphene, but has a bump at the left side for HOPG; 2) The height of 2D band peak is higher than G band peak (~1580 c $cm^{-1}$) for graphene (4 layers or less); 3) The position of G band peak moves to low wave-number from 1587 to 1581 $cm^{-1}$, when the number of graphene layers increases from one to approaching bulk HOPG; 4) The profile of D band (~1360 $cm^{-1}$) reflects the defects density, and the absence of D band corresponds to very low defects density. Based on the analysis of Raman spectra (Fig. 3), cooling rates significantly affect the amount and quality of carbon segregated from Ni surface. With a low cooling rate (0.1 °C/s), no carbon peak exists in the range of Raman shift of 1000~3000 $cm^{-1}$, indicating few carbon atoms were segregated at the surface, as the carbon atoms near surface have enough time to diffuse into the bulk of Ni substrates.. With a medium cooling rate (~10 °C/s), two prominent peaks appear at ~1583 $cm^{-1}$ and ~2704 $cm^{-1}$, corresponding to the G and 2D bands, and the higher peak intensity for 2D band relative to G band suggests that graphene films with the thickness less than 4 layers were formed. A faster cooling can reduce the migration of carbon from near the surface into the bulk, therefore enhance the carbon segregation at surface. With further increase in the cooling rate (up to ~20 °C/s), a D band at ~1360 $cm^{-1}$ in the Raman spectrum appears in addition to the G and 2D bands (Fig. 3), suggesting that although a significant amount of carbon can segregate at the surface in a short time, it may not have enough time to relax to a state with good crystallinity. These results suggest that several layers of high quality graphene can be synthesized on Ni surface with optimized medium cooling rates; while higher cooling rates result in the formation of graphite with more defects.

Transferring graphenes from metal substrates to insulators is a critical step for electronic applications. We have used silicone rubber (polymerized siloxanes) as the media to transfer 5 mm × 5 mm graphene as-grown on Ni to a glass plate. After synthesis of graphene on metal, a thin layer of silicone was applied on graphene, followed by covering with a glass plate to form a 4-layer sandwich structure (Ni/Graphene/silicon rubber/glass). After a 24-hour cure, the silicone rubber was solidified and the metal substrate was etched away with diluted $HNO_3$ solution. The transferred graphene is transparent to the eye. However, using an optical microscope with polarized light, the graphene and silicone rubber can be easily distinguished. Raman spectra confirmed that the transferred graphenes maintain their high quality. In Fig. 4, the blue curve is the Raman spectrum acquired from the graphene segregated on Ni surface with 10 °C/s cooling rate, the red curve is from the silicone rubber (with two peaks around 2900~3000 $cm^{-1}$), and the black curve is from transferred graphene. Features in the spectrum of pre-transfer graphene are maintained in the spectrum of the post-transfer graphene. Silicone peaks can be seen from the transferred graphene on the glass plate, probably because the graphene layer is so thin that the signal from silicone rubber can pass through it and be detected.

We also investigated effects of $H_2$ in growth atmosphere and roughness of metal substrates on the uniformity of graphene layers. With high volume $H_2$ annealing for one hour before introducing hydrocarbon gases, as is the case for the data presented in this paper, the uniformity of graphenes is significantly enhanced. The function of $H_2$ is believed to eliminate some impurities in metal substrates, such as S and P, which may cause the local variation of the carbon dissolvability in the metal substrates.[16] In addition, atomic H can help etch away defects in carbon (with dangling bond) at elevated temperatures. We also found that more uniform and thinner graphenes were synthesized on the smoother Ni substrates.

In summary, we have synthesized several layers of graphene on Ni substrates by a surface segregation process and transferred them to glass substrates. The graphene films keep their high quality with the usage of $HNO_3$ and mechanical operation in TEM sample preparation and film transfer, as confirmed by TEM and Raman data. Cooling rates significantly affect the thickness of graphene and the amount of defects, and the quality of graphene films can be controlled by varying the growth conditions (e.g., $H_2$ in the growth atmosphere) and surface roughness of the substrates. These results indicate that the surface segregation from metals in ambient pressure with controlled cooling rates could offer a high quality and low cost synthesis approach for graphene electronics.

ACKNOWLEDGMENT

This work is supported by the CAM Special Funding at University of Houston.

[1]A. K. Geim and K. S. Novoselov, Nat. Mater. **6**, 183 (2007).

[2]K. S. Novoselov, A. K. Geim, S. V. Morozov, D. Jiang, Y. Zhang, S. V. Dubonos, I. V. Grigorieva, and A. A. Firsov, Science **306**, 666 (2004).

[3]X. Liang, Z. Fu, and S. Y. Chou, Nano Lett. **7**, 3840 (20007).

[4]T. Ohta, A. Bostwick, T. Seyller, K. Horn, and E. Rotenberg, Science **313**, 951 (2006).

[5]C. Berger, Z. M. Song, X. B. Li, X. S. Wu, N. Brown, C. Naud, D. Mayo, T. B. Li, J. Hass, A. N. Marchenkov, E. H. Conrad, P. N. First, and W. A. de Heer, Science **312**, 1191 (2006).

[6]C. Oshima and A. Nagashima, J. Phys.: Condens. Matter **9**, 1 (1997).

[7]A. N. Obraztsov, E. A. Obraztsova, A. V. Tyurnina, and A. A. Zolotukhin, Carbon **45**, 2017 (2007).

[8]C.Gomez-navarro, R.T.Weitz, A.M.Bitter, M.Scolari, A.Mews, M.Burghard, K.Kern, Nano Lett., **7**, 3499 (2007).

[9]S. Gilje, S. Han, M. Wang, K. L. Wang, R. B. Kaner, Nano Lett. **7**, 3394 (2007).

[10]X. L. Li, X. R. Wang, L. Zhang, S. W. Lee, and H. J. Dai, Science **319**, 1229 (2008).

[11]A. Hayes and J. Chipman, Trans. Am. Inst. Min. Metall. Eng. **135**, 85 (1939).

[12]A. Y. Tontegode, Prog. Surf. Sci. **38**, 201 (1991).

[13]M. Thuvander and H.-O. Andrén, Mater. Charact. **44**, 87 (2000).

[14]A. Gupta, G. Chen, P. Joshi, S. Tadigadapa, and P. C. Eklund, Nano Lett. **6**, 2667 (2006).

[15]A. C. Ferrari, J. C. Meyer, V. Scardaci, C. Casiraghi, M. Lazzeri, F. Mauri, S. Piscanec, D. Jiang, K. S. Novoselov, S. Roth, and A. K. Geim, Phys. Rev. Lett. **97**, 187401 (2006).

[16]H. H. Angermann and Z. Horz, Appl. Surf. Sci. **70/71**, 163 (1993).

FIGURE CAPTIONS

**Figure 1.** Illustration of carbon segregation at metal surface.

**Figure 2.** TEM images of graphenes. A) Low magnification image with step shaped edges, highlighted by red dash lines. Inset shows an SAED pattern of the graphene film. B) HRTEM image of wrinkles in the graphene film, apparently of 3-4 layers.

**Figure 3.** Raman spectra of segregated carbon at Ni surface with different cooling rates.

**Figure 4.** Raman spectra of graphene before and after transferring from Ni to glass.

hydrocarbon gas     metal

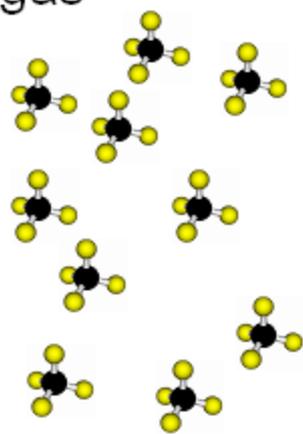
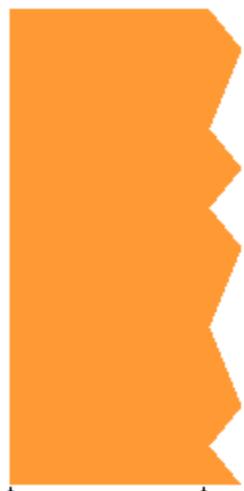

carbon
dissolving

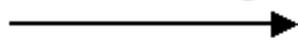

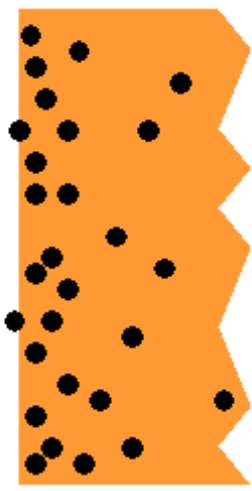

surface     body

extremely fast cooling          Fast/medium cooling          slow cooling

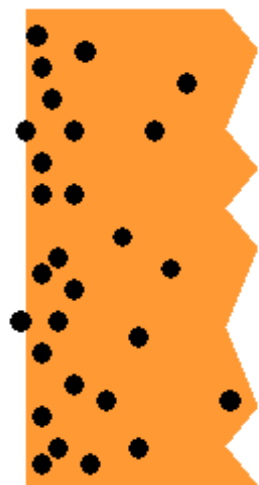
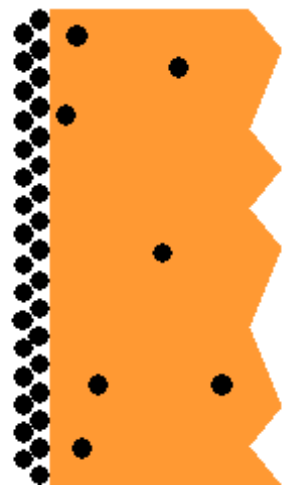
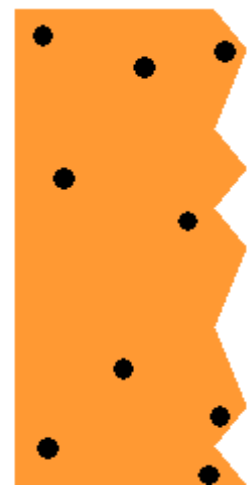

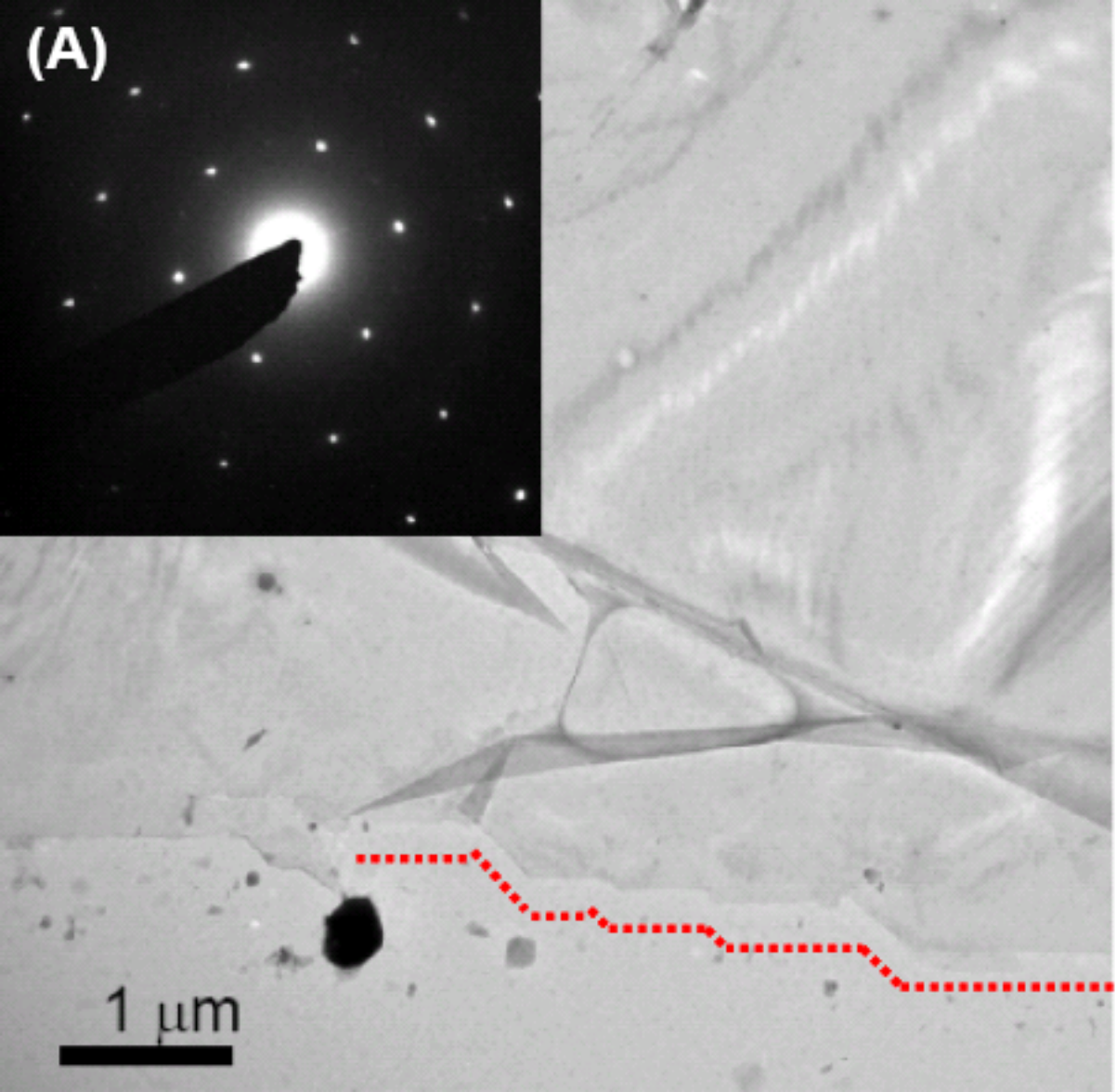

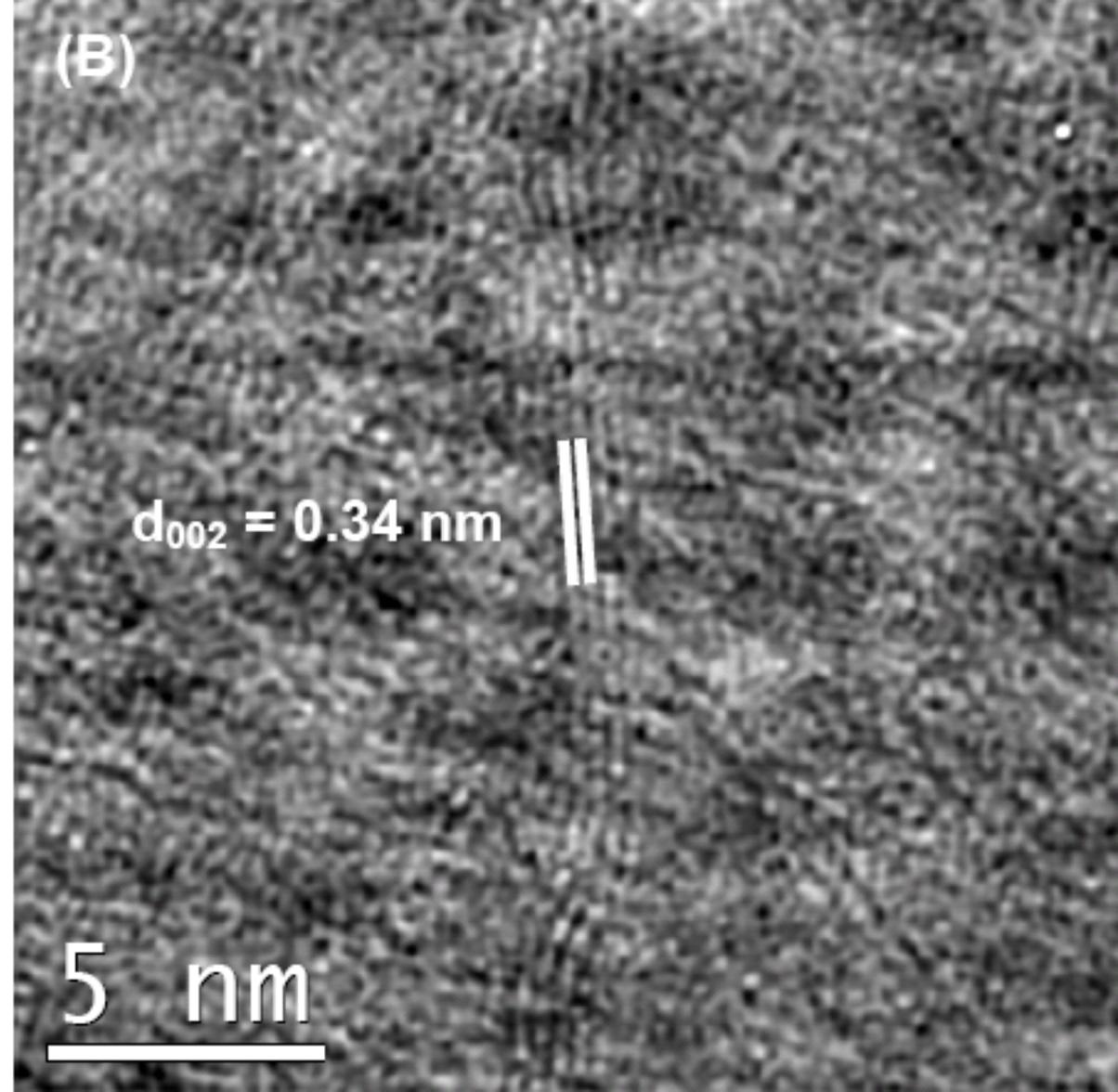

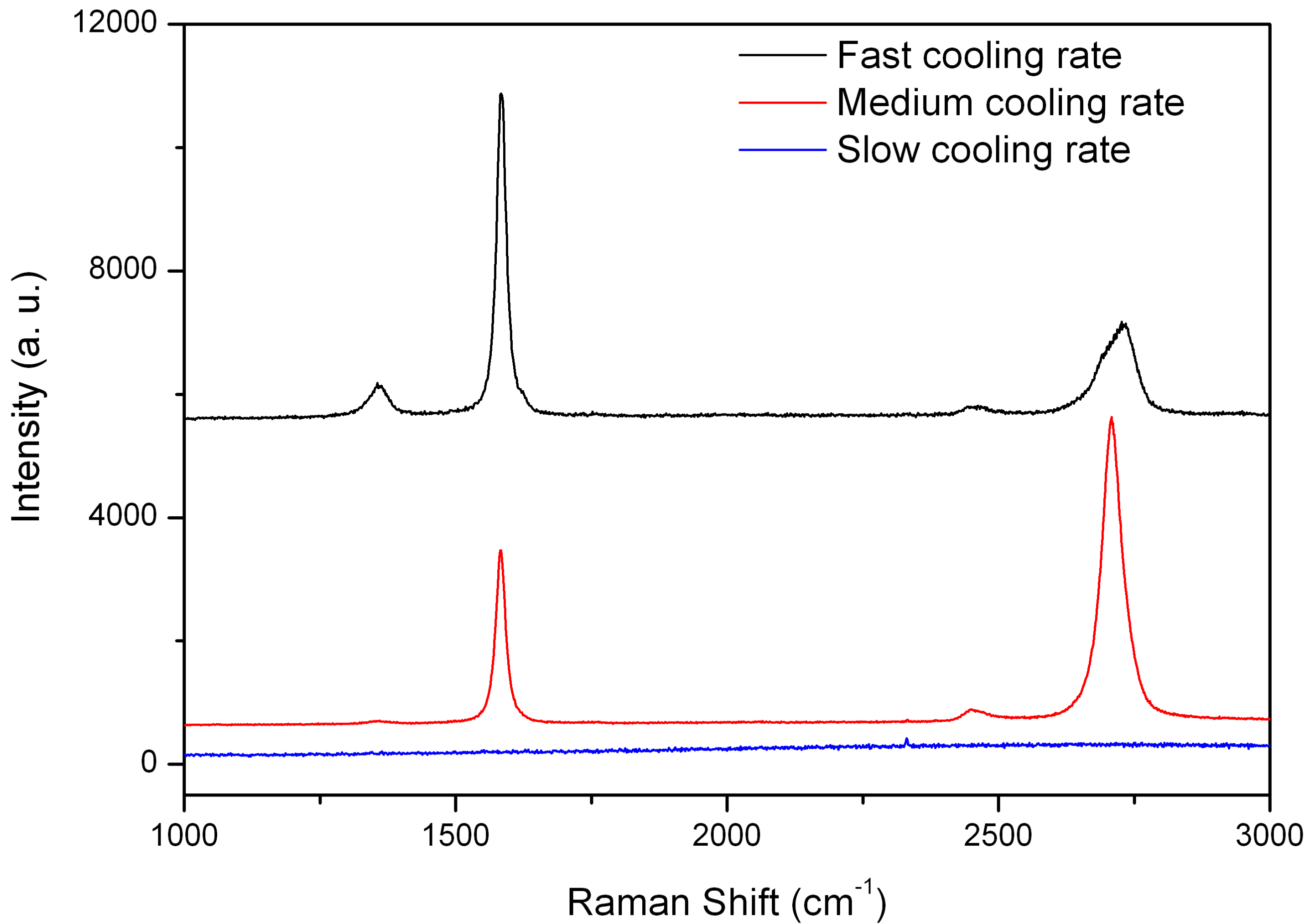

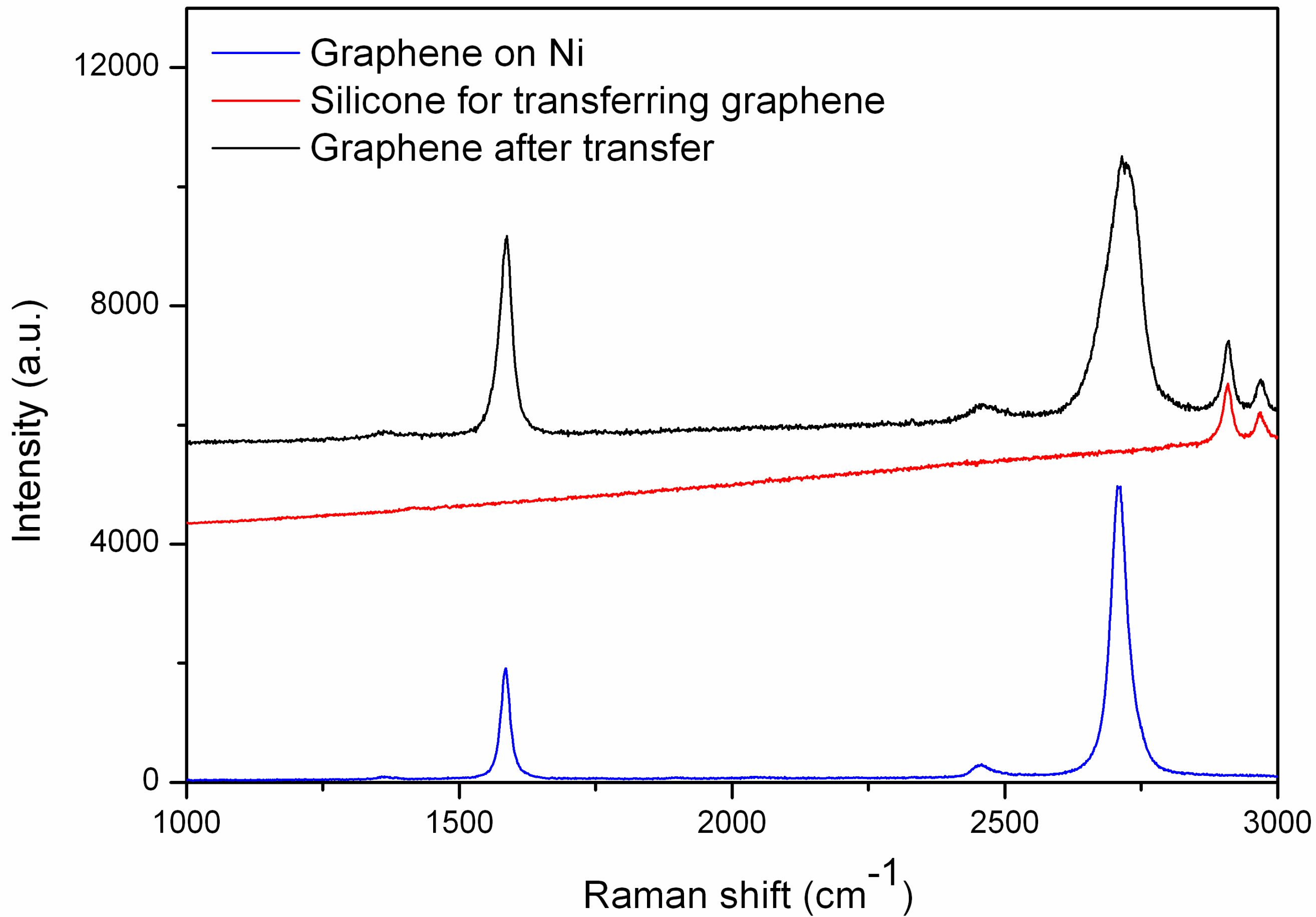